# Preferential Localization of the Bacterial Nucleoid

**Marc Joyeux**

Laboratoire Interdisciplinaire de Physique, CNRS and Université Grenoble Alpes, 38400 Grenoble, France; marc.joyeux@univ-grenoble-alpes.fr

**Abstract:** Prokaryotes do not make use of a nucleus membrane to segregate their genetic material from the cytoplasm, so that their nucleoid is potentially free to explore the whole volume of the cell. Nonetheless, high resolution images of bacteria with very compact nucleoids show that such spherical nucleoids are invariably positioned at the center of mononucleoid cells. The present work aims to determine whether such preferential localization results from generic (entropic) interactions between the nucleoid and the cell membrane or instead requires some specific mechanism, like the tethering of DNA at mid-cell or periodic fluctuations of the concentration gradient of given chemical species. To this end, we performed numerical simulations using a coarse-grained model based on the assumption that the formation of the nucleoid results from a segregative phase separation mechanism driven by the de-mixing of the DNA and non-binding globular macromolecules. These simulations show that the abrupt compaction of the DNA coil, which takes place at large crowder density, close to the jamming threshold, is accompanied by the re-localization of the DNA coil close to the regions of the bounding wall with the largest curvature, like the hemispherical caps of rod-like cells, as if the DNA coil were suddenly acquiring the localization properties of a solid sphere. This work therefore supports the hypothesis that the localization of compact nucleoids at regular cell positions involves either some anchoring of the DNA to the cell membrane or some dynamical localization mechanism.

---

## 1. Introduction

This work theoretically addresses the question of the preferential localization of the nucleoid inside bacterial cells. The nucleoid is the region of prokaryotic cells that contains their genomic DNA [1,2]. In contrast with eukaryotic cells, this DNA is not separated from the rest of the cytosol by a bounding membrane, but the nucleoid nevertheless occupies only a fraction of the cell. The nucleoid is a highly dynamical entity, with sizes that fluctuate sensitively with several factors, like the richness of the nutrients [3–7], the cell cycle step [8,9], and the concentration of antibiotics [5–7,10–14]. The question of the preferential localization of the nucleoid arises as soon as it is significantly smaller than the cell, which is the case, for example, for cells growing in rich media [3–7] and cells treated with chloramphenicol [5–7,11–14]. The point is to determine whether, under such circumstances, unconstrained nucleoids tend to localize preferentially close to the cell membrane or deep inside the cell, and why this is so. Theory indicates that a large solid sphere immersed in a sea of smaller ones localizes preferentially close to the bounding walls [15–20]. Moreover, when these walls are not planar, the sphere is most often found in the regions with the smallest radius of curvature [16,19]. Consequently, if the nucleoid and the surrounding macromolecules do behave like solids (with regard to localization) and are not subject to any constraints (except for confinement), then the nucleoid is expected to localize close to the cell membrane, and more precisely, in the hemispherical caps of rod-like bacteria (bacilli). It has however long been known that compact nucleoids are preferentially positioned at the center of mononucleoid cells and at one quarter and three quarters of the length of binucleoid cells [21–24], and these early observations have been confirmed by several recent high-resolution experiments [5,6,10,12–14,25,26]. A possible cause for this discrepancy between theoretical predictions and experimental observations may of course simply be that the nucleoid and the surrounding macromolecules do not behave like solids. More precisely, it is known that approximately 30% of the mass of the nucleoid consists of RNA and various nucleoid associated proteins, which can penetrate and diffuse inside DNA loops [2,27]. The nucleoid is consequently at best a soft solid, and it has recently been suggested that softness might trigger inner cavity positioning of solid spheres [28]. Moreover, recent simulations based on a coarse-grained model, where DNA is represented as a hyperbranched hard-sphere polymer and crowding macromolecules as freely jointed chains of hard spheres, display segregation of the DNA in the central part of the nucleoid, and consequently also support the hypothesis that the nucleoid does not behave like a solid sphere [29]. However, a second plausible cause for the discrepancy between theoretical predictions and experimental observations may be that the nucleoid is perhaps not unconstrained, but is instead tethered to the cell membrane [2,30]. It is indeed known that protein complexes involving PopZ, RacA-divIVA, and HubP help tether the origin of replication of, respectively, *C. crescentus*, sporulating *B. subtilis*, and the large chromosome of *V. cholerae*, close to the cell poles [31–33]. Moreover, certain transcription factors anchor, at least temporarily, their cognate binding site to the cell membrane [34–36], and formation of a complex involving DNA, the Noc protein, and the cellular membrane ensures that the septum does not form over the nucleoid [37]. It is therefore reasonable to speculate that the observed preferential positioning of compact nucleoids at regular cell positions may involve some kind of anchoring of the DNA molecule to the cell membrane, presumably close to the center of mononucleoid cells and at one quarter and three quarters of the length of binucleoid cells. Finally, it is well-known that the periodic oscillations of the concentrations of the components of the Min system are responsible for the assembly of the Z-ring at mid-cell [38], and the regular positioning of compact nucleoids may be dynamically regulated via a similar mechanism.

The present work aims precisely at shedding light on this question and at determining whether the localization of compact nucleoids at regular positions in bacterial cells results from straightforward interactions between the nucleoid and the cell membrane or instead requires some kind of specific mechanism, like tethering or periodic fluctuations of concentration gradients.

Let us here emphasize that the nature of the mechanism, which is responsible for the 100- to 1000-fold compaction of the DNA inside the nucleoid compared to its expected volume in standard saline solutions (estimated for example from the Worm Like Chain model [39]), is itself a longstanding, but still lively debated question [40–44]. It is becoming increasingly clear that the formation of plectonemes, the bridging of DNA duplexes by nucleoid proteins, and the action of short-range attractive forces, which are commonly evoked to rationalize the formation of the nucleoid, actually have a rather moderate effect on its size (see for example Ref. [43] and references therein). In contrast, the proposition that non-binding globular macromolecules may be able to compact the genomic DNA strongly but gradually [45–49] has recently received strong support, both from experiments [50–52] and computational simulations [53–60]. The overall repulsion between all components of the system can indeed induce de-mixing of the DNA and the other macromolecules, which eventually leads to the separation of the cytoplasm into two phases [61], one of them being rich in DNA and poor in the other macromolecules (the nucleoid), and the other one being essentially composed of crowders and almost deprived of DNA (the rest of the cytosol).

The hypothesis that the formation of the nucleoid is triggered by repulsive interactions between the DNA and macromolecular crowders is actually the starting point of the present work, which elaborates on a coarse-grained model that was proposed recently to investigate this question [57,58]. According to simulations performed with this beads-and-strings model, where each DNA bead represents 15 base pairs, compaction of long DNA molecules by non-binding spherical crowders is governed by the effective volume fraction of the crowders and increases sharply up to nucleoid-like values slightly below the jamming threshold [57,58], where the components of the system cease to flow [62]. These previous studies moreover established that, in poly-disperse systems, the largest crowders de-mix preferentially from the DNA [57,58]. Quite importantly, in these studies, the center of the confinement sphere (representing the cell membrane) was repositioned on top of the center of mass of the DNA chain after each time step, in order to make sure that the results were not affected by the slow dynamics of the compact nucleoid inside the cell and its eventual sticking to the bounding wall for significant amounts of time. In the present work, the centering step is instead omitted and the analysis focuses precisely on the motion and preferential localization of the nucleoid for different crowder concentrations and cell shapes. The goal is to understand whether purely entropic forces are, by themselves, capable of driving the regular positioning of the compact nucleoid inside the cell or if unconstrained nucleoids instead localize close to bounding walls like solid spheres and some specific mechanism is needed to achieve regular inner positioning.

Before describing the model and discussing the results obtained therewith, it may be worth emphasizing that the question addressed in this work is quite difficult to handle from a purely theoretical point of view. First, the radius of gyration of the polymer coil (DNA) is much larger than the diameter of the spherical crowders, which implies that the bacterial nucleoid pertains to the so-called "protein limit", for which the theoretical understanding [63–65] is much less satisfactory than for the opposite "colloid limit" [66–68]. In particular, phase diagrams are rather difficult to predict in this limit [69–72]. One of the major difficulties is that the deformations of the semi-rigid polymer coil, which enable the DNA polymer to fit into void spaces between neighboring crowders, must be explicitly taken into account [72,73]. Moreover, neither the crowders nor the DNA can be considered as dilute or semi-dilute when the bacterial DNA is compacted inside the nucleoid, which precludes the use of the corresponding simplifications in the theoretical treatment [73–75]. Similarly, DNA–DNA self-avoidance cannot be neglected, so that the DNA cannot be considered as an ideal polymer [76]. Finally, it is known that many-body interaction effects have to be taken into account to describe mixtures of spherical crowders and polymers in the protein limit [72,76]. These terms are particularly crucial for the compaction of a long DNA molecule upon an increase of the number of spherical crowders, as is, for example, illustrated by the fact that the DNA does not coalesce inside a compact nucleoid when it is cut into many shorter pieces. While certainly of less general purpose than analytical predictions, the model described below and the numerical investigation thereof allow one to cope adequately with all of these theoretical difficulties and get a tentative answer to the question addressed in this work.

## 2. Materials and Methods

The model used in this study is close to the model proposed recently to investigate the compaction of DNA by non-binding globular crowders [57]. It also shares several common points with those proposed previously to investigate facilitated diffusion [77–79], as well as the interaction of H-NS proteins and DNA [80–82]. The model is described briefly in this section for the sake of clarity and completeness.

Genomic DNA is modeled as a circular chain of $n = 1440$ beads of radius $a = 1.78$ nm separated at equilibrium by a distance $l_0 = 5.0$ nm, where each bead represents 15 consecutive base pairs. The DNA chain is enclosed in a confinement chamber, which volume is chosen so that the concentration of nucleotides is close to physiological values (around 10 mM), in spite of the 200-fold reduction relative to the length of the DNA of *E. coli* cells. Simulations were run with two different geometries of the confinement chamber, namely: (i) A sphere of radius $R_0$, and (ii) a cylinder of radius $R_0$ and length $R_0$ capped at each end by a hemisphere of radius $R_0$, which model the rigid walls of cocci and bacilli, respectively. Choosing $R_0 = 120$ nm for the sphere and $R_0 = (2/5)^{1/3} \times 120 \approx 88.42$ nm for the capped cylinder leads to an identical volume $V$ for the two confinement chambers. In addition to the DNA chain, $N = 1500$, 1750, or 2000 spherical crowders of radius $b = 6.5$ nm are also enclosed in the confinement chamber.

The potential energy of the system is written as the sum of four terms (Equation (1))

$$E_{\text{pot}} = V_{\text{DNA}} + V_{\text{DNA/C}} + V_{\text{C/C}} + V_{\text{wall}}, \tag{1}$$

which describe the internal energy of the DNA chain, DNA/crowder interactions, crowder/crowder interactions, and repulsion exerted by the bounding walls, respectively.

The internal energy of the DNA chain is further expanded as the sum of 3 contributions (Equation (2))

$$V_{\text{DNA}} = \frac{h}{2}\sum_{k=1}^{n}(l_k - l_0)^2 + \frac{g}{2}\sum_{k=1}^{n}\theta_k^2 + e_{\text{DNA}}^2\sum_{k=1}^{n-2}\sum_{j=k+2}^{n}H(\|\mathbf{r}_k - \mathbf{r}_j\|), \tag{2}$$

where (Equation (3))

$$H(r) = \frac{1}{4\pi\varepsilon\, r}\exp\left(-\frac{r}{r_D}\right), \tag{3}$$

which describe the stretching, bending, and electrostatic energy of the DNA chain, respectively. $\mathbf{r}_k$ denotes the position of the center of DNA bead $k$, $l_k$ the distance between the centers of two successive beads, and $\theta_k$ the angle formed by the centers of three successive beads. The stretching term aimed to avoid a rigid rod description and has no biological meaning. Setting $h$ to 1000 $k_{\text{B}}T/l_0^2$, where $T = 298$ K is the temperature of the system, ensures that $|l_k - l_0|$ remains, on average, of the order of 0.02 $l_0$, in spite of the compression forces exerted by the crowders. The bending rigidity constant $g$ was set to 9.82 $k_{\text{B}}T$, in order to get the correct persistence length for DNA, $\xi = gl_0/(k_{\text{B}}T) \approx 49$ nm [83]. Finally, the electrostatic repulsion between DNA beads is written as a sum of Debye-Hückel potentials [84], where $e_{\text{DNA}} = -12.15\,\bar{e}$ denotes the value of the point charge placed at the center of each DNA bead, $\varepsilon = 80\,\varepsilon_0$ is the dielectric constant of the medium, and $r_D = 1.07$ nm the Debye length inside the medium, which corresponds to a concentration of monovalent salts close to 100 mM, as is often assumed in bacteria. $e_{\text{DNA}}$ is significantly smaller than the total charge carried by the phosphate groups of 15 base pairs ($-30\,\bar{e}$), in order to account for counter-ion condensation [85,86]. Note that $l_0$ is too large compared to $r_D$ to warrant that different parts of the DNA chain will never cross, but such crossings are rather infrequent and appear to affect the geometry of the DNA chain only to a limited extent. Finally, electrostatic interactions between nearest-neighbors are not included in Equation (2), because it is considered that they are already accounted for in the stretching and bending terms.

DNA/crowder and crowder/crowder interactions are expressed as sums of Debye-Hückel potentials with hard core (Equation (4))

$$\begin{aligned} V_{\text{DNA/C}} &= e_{\text{DNA}}\, e_{\text{C}} \sum_{k=1}^{n}\sum_{K=1}^{N} H(\|\mathbf{r}_k - \mathbf{R}_K\| - b) \\ V_{\text{C/C}} &= e_{\text{C}}^2 \sum_{K=1}^{N-1}\sum_{L=K+1}^{N} H(\|\mathbf{R}_K - \mathbf{R}_L\| - 2b)\,, \end{aligned} \tag{4}$$

where $\mathbf{R}_K$ denotes the position of the center of crowding sphere $K$ and $e_{\text{C}}$ the electrostatic charge placed there. As in previous work, $e_{\text{C}}$ was set to $e_{\text{C}} = e_{\text{DNA}}$, which ensures that strong compaction of the DNA chain is obtained for an effective volume fraction of the crowders (Equation (5))

$$\rho = \frac{4\pi N(b + \Delta b)^3}{3\,V} \tag{5}$$

in the range $0.60 \leq \rho \leq 0.70$, that is close to the jamming threshold for hard spheres [57,58]. In Equation (5), $b + \Delta b = 8.3$ nm represents the effective radius of the crowding spheres, that is half the distance at which the electrostatic repulsion energy is equal to thermal energy. Simulations performed with $N = 1500$, 1750, and 2000, crowders correspond to $\rho = 0.50$, 0.58, and 0.66, respectively.

Finally, the repulsion exerted by the confinement chamber is written in the form (Equation (6))

$$V_{\text{wall}} = \zeta \sum_k \left( \left(1 + \frac{d_k}{R_0}\right)^6 - 1 \right) + \zeta \sum_K \left( \left(1 + \frac{D_K}{R_0}\right)^6 - 1 \right), \tag{6}$$

where the sums extend only to the particles that have trespassed the limits of the confinement chamber, $d_k$ (respectively, $D_k$) denotes the distance from the center of such DNA (respectively, crowder) particles to the bounding wall, and $\zeta = 1000\,k_{\text{B}}T$.

The dynamics of the system were investigated by numerically integrating overdamped Langevin equations. Practically, the updated positions at time step $i$+1 were computed from the positions at time step $i$ according to (Equation (7))

$$\begin{aligned} \mathbf{r}_k^{(i+1)} &= \mathbf{r}_k^{(i)} + \frac{\Delta t}{6\pi\eta\, a}\mathbf{f}_k^{(i)} + \sqrt{\frac{2\,k_B T\,\Delta t}{6\pi\eta\, a}}\, x_k^{(i)} \\ \mathbf{R}_K^{(i+1)} &= \mathbf{R}_K^{(i)} + \frac{\Delta t}{6\pi\eta\, b}\mathbf{F}_K^{(i)} + \sqrt{\frac{2\,k_B T\,\Delta t}{6\pi\eta\, b}}\, X_K^{(i)}\,, \end{aligned} \tag{7}$$

where $\Delta t = 20$ ps is the integration time step, $\mathbf{f}_k^{(i)}$ and $\mathbf{F}_K^{(i)}$ are vectors of inter-particle forces arising from the potential energy $E_{\text{pot}}$, $x_k^{(i)}$ and $X_K^{(i)}$ are vectors of random numbers extracted from a Gaussian distribution of mean 0 and variance 1, $T = 298$ K is the temperature of the system, and $\eta = 0.00089$ Pa s is the viscosity of the buffer. Each trajectory was integrated for a total time duration ranging from 60 to 200 ms, depending on the geometry of the confinement chamber and the number of crowders. The first 10 to 50 ms of each trajectory were used for equilibration and the remainder for production. Representative snapshots of such trajectories are shown in Figures 1 and 2. We note in passing that the time scales reported in the present work, which are in the range 1–100 ms, are by no means representative of the corresponding time scales in real cells, because of (i) the coarse-grained nature of the model, which accelerates the dynamics of the system by two to three orders of magnitude compared to all-atom representations, and (ii) the fact that the model represents only ≈1/200 of an *E. coli* cell, while the relaxation times of long polymers increase rapidly with their length $n$.

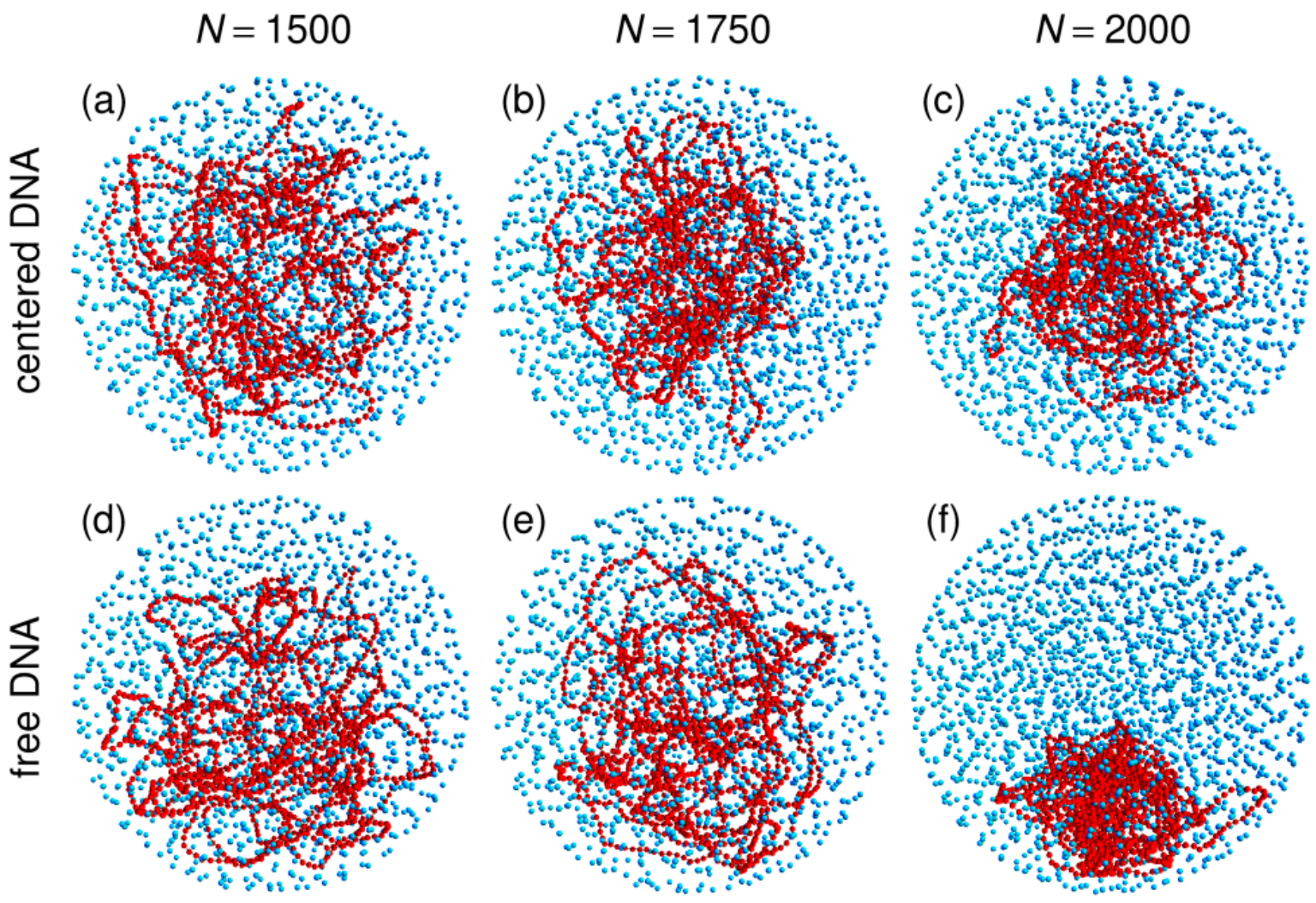

**Figure 1.** Representative snapshots of simulations performed with the spherical confinement chamber and $N = 1500$ (panels (**a**) and (**d**)), 1750 (panels (**b**) and (**e**)), or 2000 (panels (**c**) and (**f**)) crowders. For panels (**a**), (**b**), and (**c**), the center of the confinement sphere was repositioned on top of the center of mass of the DNA chain after each integration time step, while the centering step was omitted for panels (**d**), (**e**), and (**f**). DNA beads are colored in red and spherical crowders in cyan. Crowders are represented at $\frac{1}{4}$ of their actual radius, in order that the DNA chain may be seen through the layers of crowders. The confinement sphere is not shown.

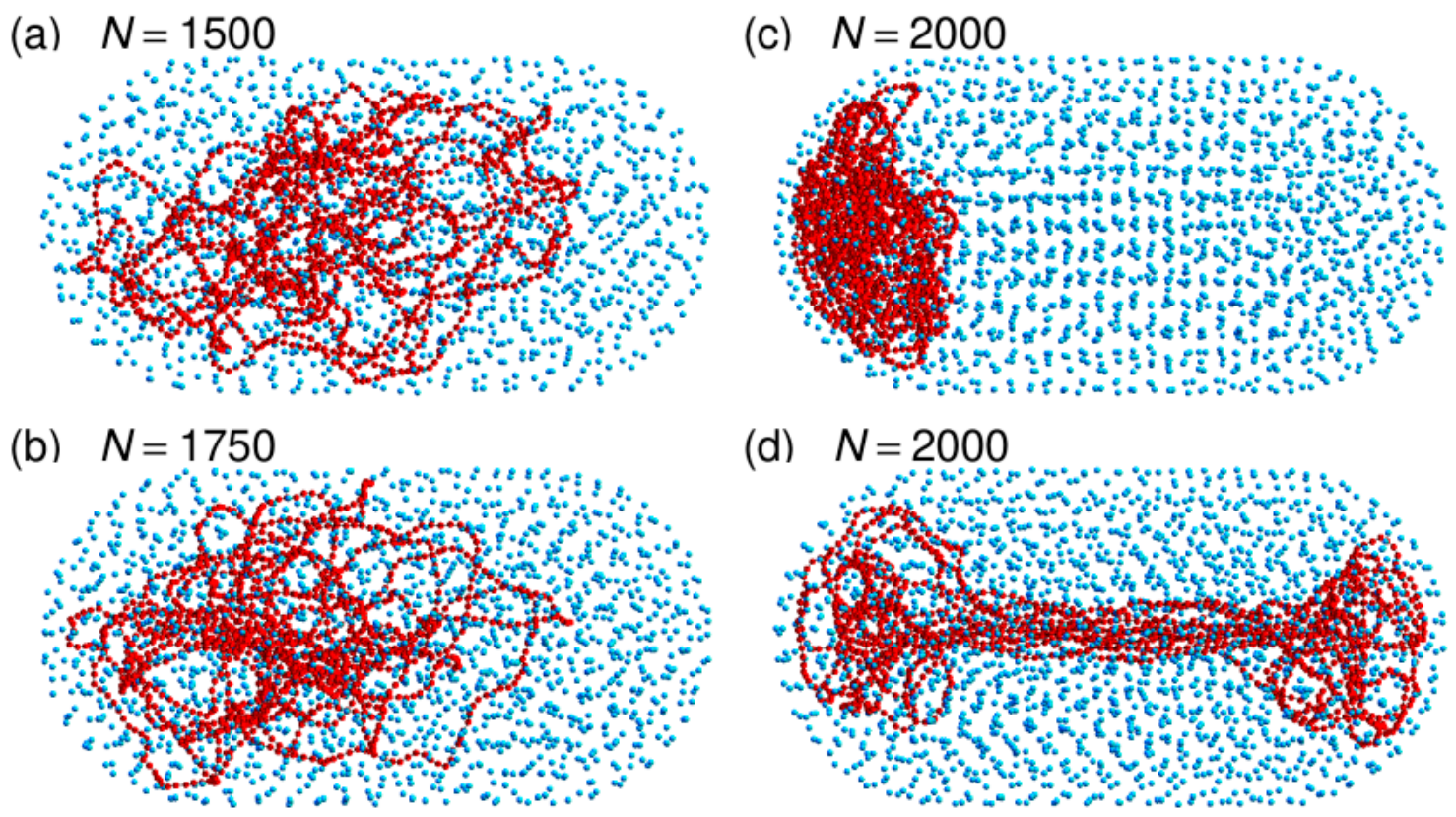

**Figure 2.** Representative snapshots of simulations performed with the capped confinement cylinder and $N = 1500$ (panel (**a**)), 1750 (panel (**b**)), or 2000 (panels (**c**) and (**d**)) crowders. These simulations were run without repositioning the center of the confinement sphere on top of the center of mass of the DNA chain. DNA beads are colored in red and spherical crowders in cyan. Crowders are represented at $\frac{1}{4}$ of their actual radius, in order that the DNA chain may be seen through the layers of crowders. The capped confinement cylinder is not shown.

## 3. Results

### *3.1. Compaction of DNA Coils with Increasing Crowder Density*

Previous work investigated the behavior of a circular chain enclosed in a spherical confinement chamber, together with an increasing number $N$ of crowding particles, under the constraint that the center of the confinement sphere is repositioned on top of the center of mass of the circular chain, $\mathbf{r}_{\mathrm{CM}}$, after each integration step [57,58]. It was shown that the average size of the coil decreases with increasing values of $N$ and that compaction is driven by $\rho$, the effective volume fraction of the crowders defined in Equation (5). Moreover, compaction increases sharply close to the jamming transition at $\rho \approx 0.65$ [57,58]. These results are illustrated on the top row of Figure 1, which shows representative snapshots of trajectories run with $N = 1500$, 1750, and 2000 crowders (corresponding to $\rho = 0.50$, 0.58, and 0.66, respectively), and the inset of Figure 3, which shows the evolution with $N$ of $\langle R_{\mathrm{coil}} \rangle$, the mean value of the average radius of the coil (Equation (8))

$$R_{\mathrm{coil}} = \frac{1}{n} \sum_{k=1}^{n} \| \mathbf{r}_k - \mathbf{r}_{\mathrm{CM}} \| . \quad (8)$$

Simulations were performed by first letting the circular chain equilibrate in the confinement sphere, which it fills almost homogeneously. Mean radius of the DNA coil under these conditions is $\langle R_{\mathrm{coil}} \rangle \approx 78.8$ nm. Crowders were then introduced at random, non-overlapping, and homogeneously distributed positions, and the system was allowed to equilibrate again for 20 ms (for $N = 1500$ and $N = 1750$) or 50 ms (for $N = 2000$). Equilibration time is longer for $N = 2000$ because of the strongly reduced mobility of the crowders close to the jamming threshold. Finally, $R_{\mathrm{coil}}$ was averaged for 50 ms along each trajectory. The increasing compaction of DNA coils with an increasing number of crowders is clearly seen in the inset of Figure 3. As is also illustrated in Figure 1c, compaction of the circular chain is particularly strong close to the jamming threshold, with an average radius as small as $\langle R_{\mathrm{coil}} \rangle \approx 55.0$ nm for $N = 2000$.

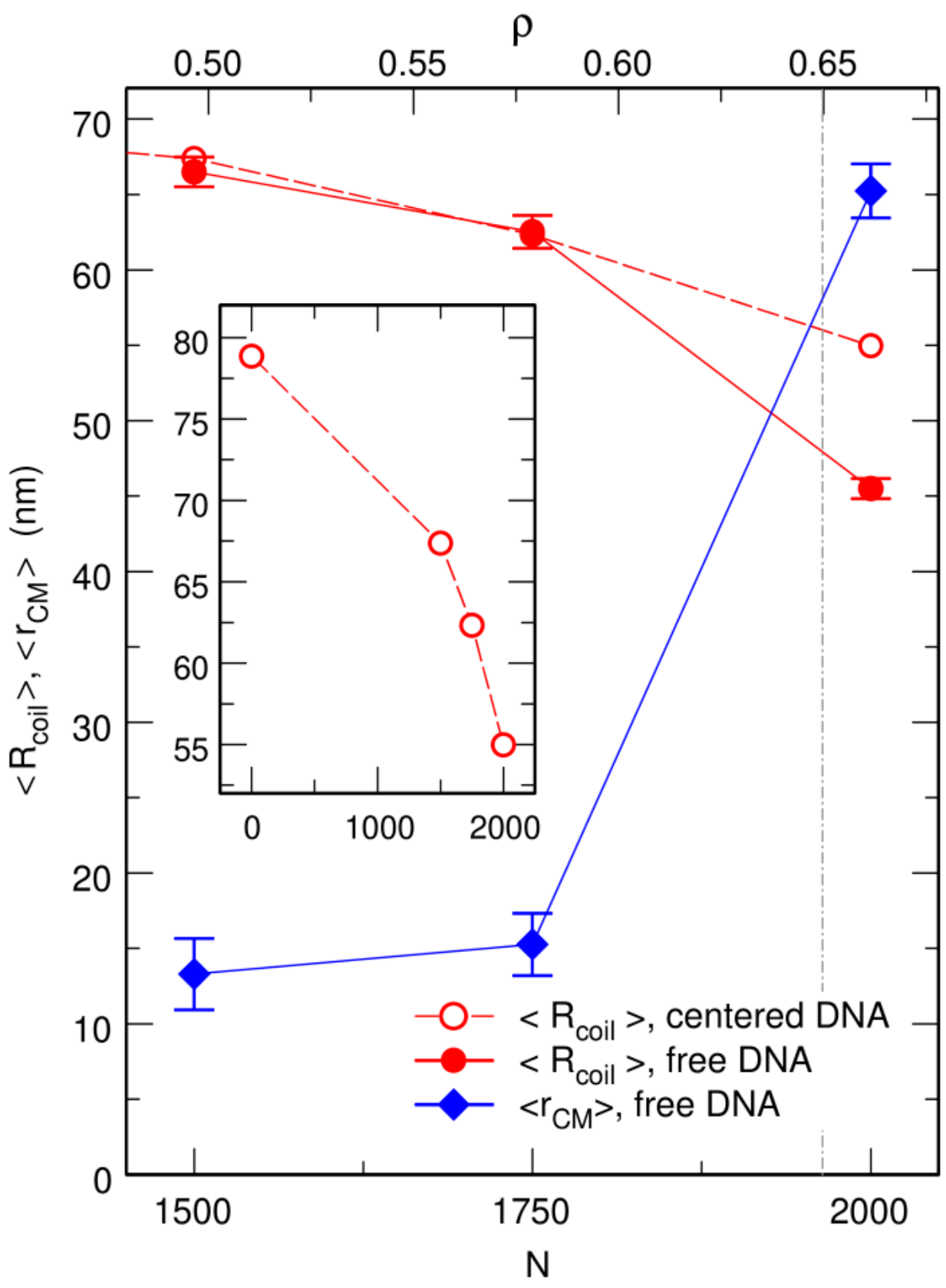

**Figure 3.** Plot, as a function of the number $N$ of spherical crowders (lower $x$ axis) or the effective crowder volume fraction $\rho$ (upper $x$ axis), of the mean radius of the DNA coil, $< R_{\text{coil}} >$ (circles), and the mean distance from its center of mass to the center of the confinement sphere, $< r_{\text{CM}} >$ (lozenges). Empty circles correspond to the case where the center of the confinement sphere was repositioned on top of the center of mass of the DNA chain after each integration time step, and filled circles to the case where the centering step was omitted. The inset shows the plot of $< R_{\text{coil}} >$ over a broader range of values of $N$, in order to highlight the abrupt decrease close to the jamming threshold. $< R_{\text{coil}} >$ and $< r_{\text{CM}} >$ are expressed in units of nm. The vertical dot-dashed line at $\rho = 0.65$ denotes the approximate location of the jamming threshold.

More detailed information is gained from the plots of $Q_X(r)$, the enrichment in species $X$ at a distance $r$ from the center of the confining sphere. $Q_X(r)$ is defined so that the mean number of particles of species $X$ with center located in a distance interval $[r, r + dr[$ from the center of the confining sphere, is equal to (Equation (9))

$$4\pi r^2 Q_X(r) \frac{n_X}{V}, \tag{9}$$

where $n_X$ is the total number of particles of species $X$ ($n_X = n = 1440$ for DNA beads and $n_X = N$ for crowders). Note that $Q_X(r) = 1$ for a homogeneously distributed species $X$. The plots of $Q_{\text{DNA}}(r)$ obtained from trajectories where the center of the confinement sphere was repositioned on top of the center of mass of the DNA chain after each integration time step are shown as dashed lines in Figure 4. These plots indicate that increasing the value of $N$ results in a strong increase of the enrichment in DNA beads close to the center of the confinement sphere. Enrichment values as large as $Q_{\text{DNA}}(0) \approx 10$ are obtained at the center of the confinement sphere for $N = 2000$. Periodic oscillations in the plot of $Q_{\text{DNA}}(r)$ are moreover observed for $N = 2000$. As can be checked in Figure 5, which displays superimposed plots of $Q_{\text{DNA}}(r)$ and $Q_C(r)$ for $N = 2000$, these oscillations arise from the fact that, close to the jamming threshold, crowders tend to arrange in regular concentric layers, while DNA beads localize preferentially in the interstices between two successive layers.

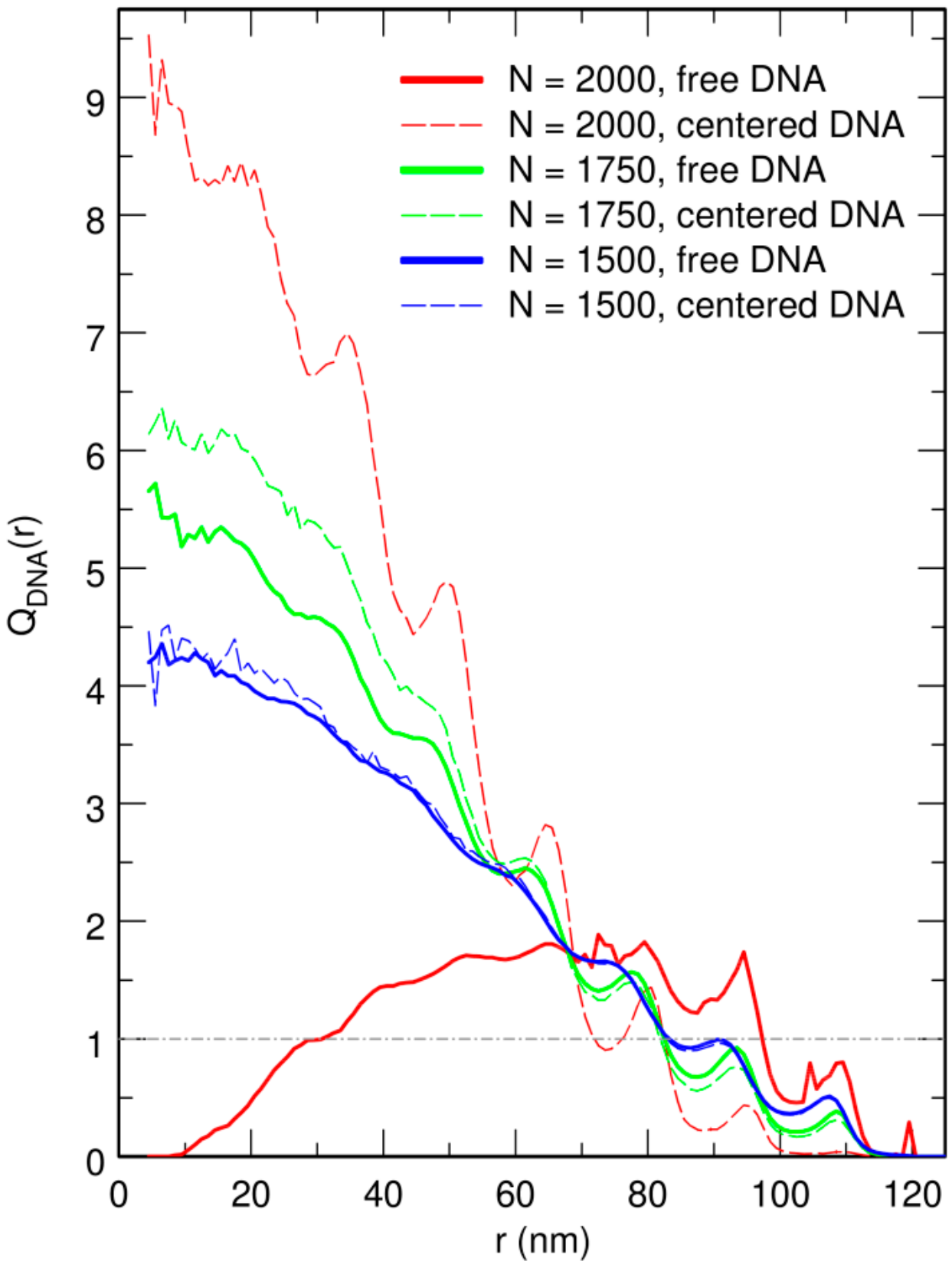

**Figure 4.** Plot of $Q_{DNA}(r)$, the enrichment in DNA beads as a function of the distance $r$ from the center of the confinement sphere, for $N = 1500$ (blue curves), 1750 (green curves), and 2000 (red curves) crowders. Dashed lines correspond to the case where the center of the confinement sphere was repositioned on top of the center of mass of the DNA chain after each integration time step, and solid lines to the case where the centering step was omitted. $Q_{DNA}(r)$ is constant and equal to 1 (dot-dashed horizontal gray line) for DNA beads distributed homogeneously all over the confinement sphere. $r$ is expressed in units of nm. $N = 1500$, 1750, and 2000 correspond to effective volume fractions of the crowders $\rho = 0.50$, 0.58, and 0.66, respectively.

### *3.2. Free DNA Coils Acquire the Localization Properties of Solid Spheres at Large Crowder Density*

One of the principal goals of this work is to understand how the conclusions summarized in the previous sub-section are affected when the center of mass of the circular chain is not constrained to remain at the center of the spherical confinement chamber. In order to answer this question, several sets of trajectories were launched, as described above, except that the centering step was omitted. Moreover, not only $\langle R_{coil} \rangle$, but also $\langle r_{CM} \rangle$ (the mean distance from the center of mass of the DNA chain to the center of the confinement sphere), were computed during the 50 ms production time windows, and results were averaged over eight different trajectories for the sake of better statistics.

Representative snapshots extracted from these simulations are shown in the bottom row of Figure 1. Visually, the vignettes for $N = 1500$ and $N = 1750$ look quite similar to those in the top row, which are extracted from simulations that included the centering step. In contrast, the vignettes for $N = 2000$ are markedly different, in that the DNA coil looks significantly more compact when the centering step is omitted and it furthermore localizes close to the bounding wall. This visual inspection is confirmed by the plots of $\langle R_{coil} \rangle$ (circles) and $\langle r_{CM} \rangle$ (lozenges) as a function of $N$, which are shown in Figure 3. While for $N = 1500$ and $N = 1750$, the values of $\langle R_{coil} \rangle$ obtained with and without the centering step almost superpose, for $N = 2000$, the mean radius of the DNA coil is about 10 nm smaller when the centering step is omitted ($\langle R_{coil} \rangle \approx 45.5$ nm instead of $\langle R_{coil} \rangle \approx 55.0$ nm). Simultaneously, the center of mass of the DNA coil, which remains on average within 15 nm from the center of the confinement sphere for $N = 1500$ and $N = 1750$, jumps to $\langle r_{CM} \rangle \approx 65.2$ nm for $N = 2000$. Examination of the plots of $Q_{DNA}(r)$ in Figure 4 leads to corroborating conclusions. Indeed, for $N = 1500$ and $N = 1750$, omission of the centering step (solid lines) results in a rather limited transfer of DNA beads towards the periphery of the confinement sphere. In contrast, for $N = 2000$, the center of the confinement sphere is totally deprived of DNA beads when the centering step is omitted, because the very compact DNA coil has migrated towards the surface of the confinement sphere and sticks to it.

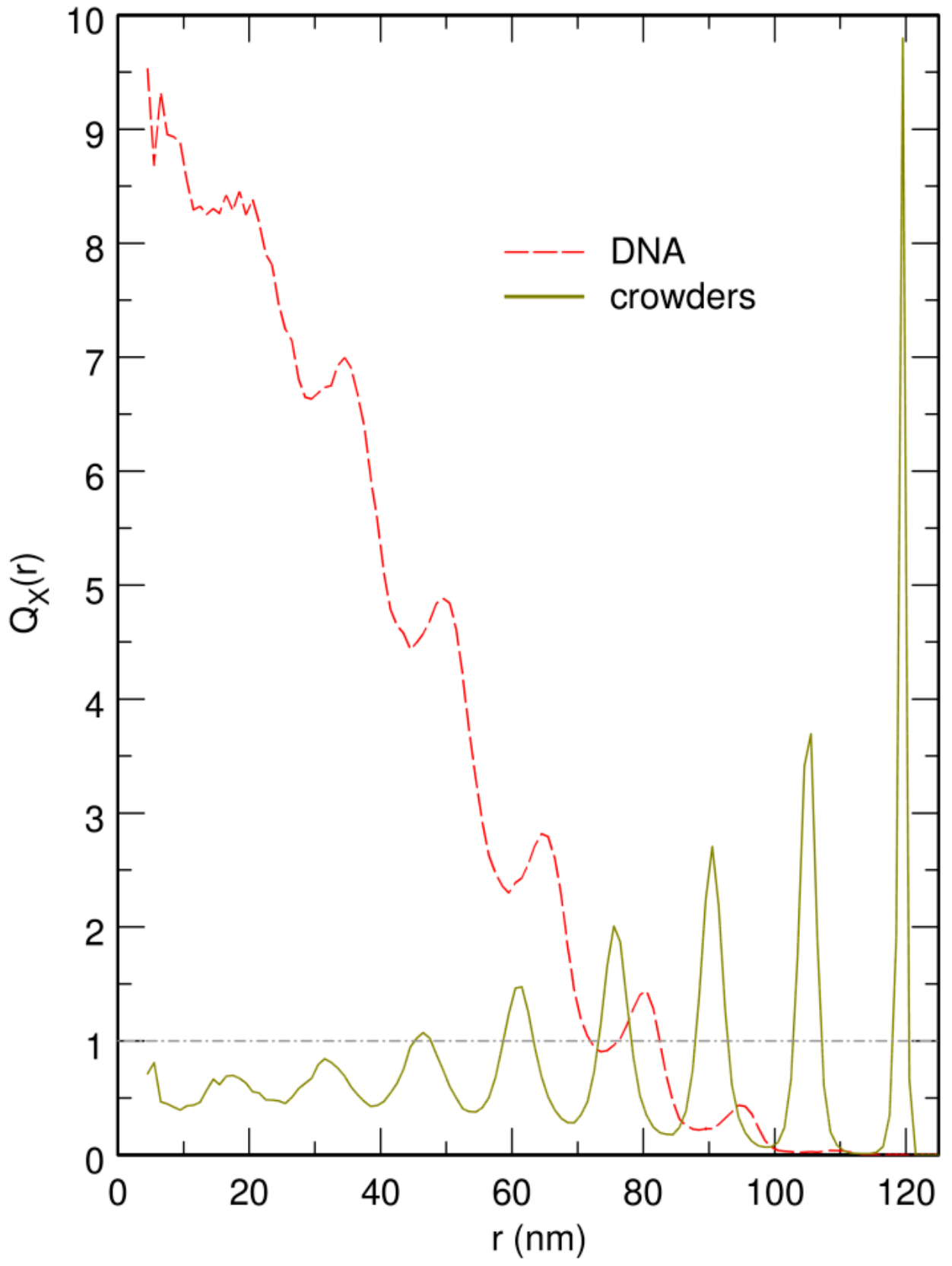

**Figure 5.** Plot of $Q_{DNA}(r)$ (red curve) and $Q_C(r)$ (brown curve), the enrichment in DNA beads and crowders, respectively, as a function of the distance $r$ from the center of the confinement sphere, for $N = 2000$ crowders. The center of the confinement sphere was repositioned on top of the center of mass of the DNA chain after each integration time step. $Q_X(r)$ is constant and equal to 1 (dot-dashed horizontal gray line) for a homogeneously distributed species X. $r$ is expressed in units of nm. $N = 2000$ corresponds to an effective volume fraction of the crowders $\rho = 0.66$, close to the jamming threshold for solid spheres.

The most straightforward explanation for this result is that free DNA coils acquire the localization properties of solid spheres at large crowder density. Indeed, for values of $\rho$ smaller than about 0.60, the DNA chain is only moderately compacted by the segregative phase separation mechanism, so that many crowders are able to diffuse inside the coil. As a consequence, most DNA beads experience almost isotropic collisions with the crowders. There is therefore no preferred positioning of the DNA coil inside the confinement sphere, which results in distributions of DNA beads with spherical symmetry and central maximum, as observed in Figure 4 for $N = 1500$ and $N = 1750$. In contrast, for values of $\rho$ close to the jamming threshold for solid spheres ($\rho \approx 0.65$), DNA/crowder de-mixing is significantly more efficient than for lower crowder density, so that much fewer crowders are able to diffuse inside the DNA coil. As a consequence, the vast majority of DNA/crowder collisions take place at the surface of the DNA coil. When diffusion brings the compact coil close to the bounding wall, the pressure exerted on the surface of the coil that faces the wall diminishes steadily and the DNA coil is pressed more and more firmly against the wall. Simultaneously, the shape of the coil adapts to some extent to that of the bounding wall, which makes the whole process even more efficient and the DNA coil is unable to detach from the wall. Except for the deformability of the DNA coil, this mechanism is identical to the one that is responsible for the fact that a big hard sphere immersed in a sea of smaller ones localizes preferentially close to bounding walls [15–20]. The results above therefore suggest that, close to the jamming threshold, the DNA coil acquires the localization properties of a solid sphere, because of the abrupt increase in DNA compaction that takes place at large crowder density. Note that we did not check whether the DNA coil simultaneously acquires other properties that are specific to solids, like for example an eventual $n^{1/3}$ dependence of $\langle R_{\text{coil}} \rangle$.

A few comments are in order here. First, it can be checked in Figures 1f and 2c,d that the DNA coil is always separated from the wall of the confinement chamber by at least one layer of crowders. This is an artifact of the model, which arises from the fact that the repulsion term in Equation (6) acts on all particles whose center (not whose periphery) trespasses the wall. Since the radius of crowding spheres is much larger than the radius of DNA beads, it is convenient from the entropic point of view that the centers of as large as possible number of crowding spheres localize on the wall, because this de facto increases the volume available to all other particles. Moreover, it may be worth emphasizing that the concentric shells of crowders, which are clearly seen in Figures 2c and 4, are due to the homogeneity of the crowders and fade progressively away with increasing crowder heterogeneity. However, as was shown previously [57,58], this does not significantly affect the compaction of the DNA coil, because the main effect of crowder heterogeneity is to let the crowders with the largest radius be expelled preferentially from the DNA coil. Such concentric shells probably do not form in real heterogeneous cytosol, but this does not invalidate the conclusions drawn here.

### 3.3. Localization of Compact DNA Coils in the Regions with Largest Boundary Curvature

As mentioned in the Introduction section, theory indicates that a big hard sphere immersed in a sea of smaller ones and enclosed in a confinement chamber with complex geometry localizes preferentially close to the regions of the bounding wall with the largest curvature [16,19]. If the conclusion drawn in the previous sub-section is correct and the DNA coil indeed acquires the localization properties of a solid sphere close to the jamming threshold, then replacing spherical confinement walls (specific to cocci) by rod-like confinement walls (specific to bacilli) should result in the migration of the coil towards the hemispherical caps of the confinement cylinder. In order to check this point, several sets of trajectories were launched with the capped confinement cylinder described in the Materials and Methods section. All trajectories with $N = 1500$ and $N = 1750$ equilibrated within less than 25 ms, while much longer equilibration time windows were again required for $N = 2000$ (see below). After the equilibration step, three quantities were averaged over time windows of 75 ms, in order to characterize the dimension and position of the DNA coil, namely $\langle R_{\text{coil}} \rangle$, $\langle h_{\text{CM}} \rangle$, and $\langle \text{abs}(x_{\text{CM}}) \rangle$, where $h_{\text{CM}}$ denotes the radial distance from the center of mass of the DNA coil to the axis of revolution of the cylinder, and $x_{\text{CM}}$ the position of the center of mass of the DNA coil along the axis of revolution of the cylinder, with $x_{\text{CM}} = 0$ at the center of the cylinder. For $N = 1500$ and $N = 1750$, results were finally averaged over eight different trajectories, while for $N = 2000$, they were averaged over only three different trajectories (see below).

Representative snapshots extracted from these simulations are shown in Figure 2 and the plots of $\langle R_{\text{coil}} \rangle$, $\langle h_{\text{CM}} \rangle$, and $\langle \text{abs}(x_{\text{CM}}) \rangle$ as a function of $N$ in Figure 6. Far from the jamming threshold, the compaction ratio of the DNA coil in the rod-like confinement chamber remains moderate and close to the value obtained for the spherical confinement chamber (for example $\langle R_{\text{coil}} \rangle = 62.5$ nm against 63.6 nm for $N = 1750$). This is not really surprising, because the two confinement chambers have the same volume $V$, so that for each value of $N$, the associated values of $\rho$ obtained from Equation (5) are identical. Moreover, the center of mass of the DNA coil remains close to the center of the confinement cylinder, both longitudinally and radially. As can be checked in Figure 6, for $N = 1750$ $\langle \text{abs}(x_{\text{CM}}) \rangle$ remains indeed as small as about 23 nm (compared to the half-length $2R_0 \approx 177$ nm of the confinement chamber) and $\langle h_{CM} \rangle$ of the order of only 8 nm (compared to the radius $R_0 \approx 88$ nm of the confinement chamber).

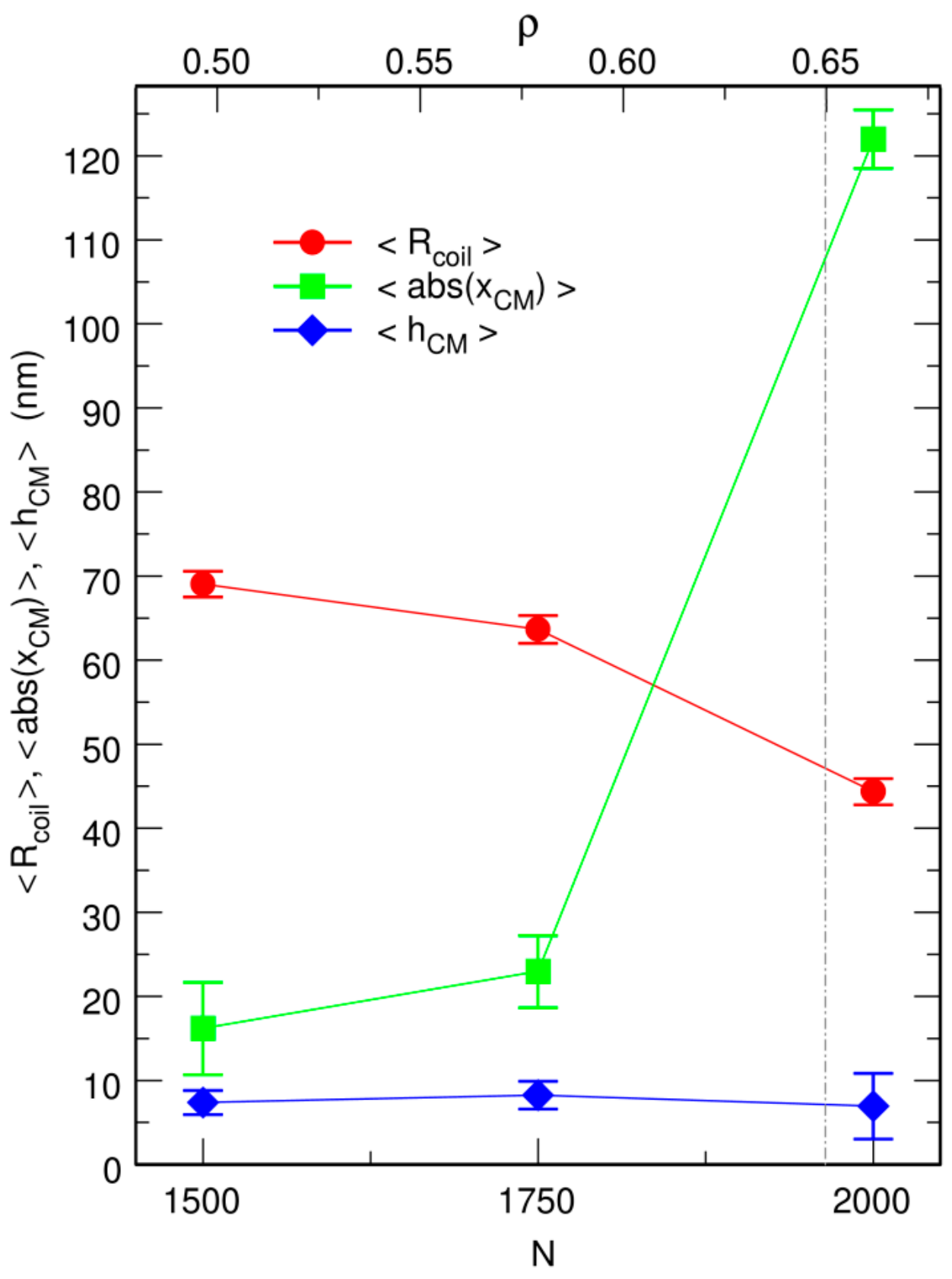

**Figure 6.** Plot, as a function of the number $N$ of spherical crowders (lower $x$ axis) or the effective crowder volume fraction $\rho$ (upper $x$ axis), of the mean radius of the DNA coil, $< R_{coil} >$ (circles), the mean radial distance from the center of mass of the DNA coil to the axis of the confinement chamber, $\langle h_{CM} \rangle$ (lozenges), and the abscissa of the center of mass of the DNA coil along the axis of the confinement chamber, $\langle \text{abs}(x_{CM}) \rangle$ (squares). These simulations were run without repositioning the center of the confinement chamber on top of the center of mass of the DNA coil. $< R_{coil} >$, $\langle h_{CM} \rangle$, and $\langle \text{abs}(x_{CM}) \rangle$ are expressed in units of nm. The vertical dot-dashed line at $\rho = 0.65$ denotes the approximate location of the jamming threshold.

As for the spherical confinement chamber, results obtained close to the jamming threshold ($N = 2000$) are however very different from those obtained at lower crowder density. Indeed, three out of the eight trajectories that were run with $N = 2000$ led within a 50 ms equilibration time window to an equilibrated conformation similar to the one shown in Figure 2c. The DNA chain is compacted to the same huge ratio as for the confinement sphere ($\langle R_{coil} \rangle = 44.4$ nm against 45.5 nm) and it fills one of the hemispherical caps of the confinement chamber. Figure 7 illustrates the sequence of events, which take place during the equilibration step and ultimately lead to this result. Within a few tens of ms, the DNA chain first forms a thick filament that elongates along the axis of the cylinder, which indicates that compaction is more rapid perpendicular to this axis than along the axis. Such a difference probably arises from the fact that the persistence length of the DNA chain (50 nm) is of the same order of magnitude as the dimensions of the confinement chamber ($R_0 \approx 88.42$ nm), so that compaction of the coil perpendicular to the largest dimension of the chamber costs significantly less bending energy than compaction along the axis of the cylinder. Compaction along the axis of the cylinder does however continue and the DNA chain progressively acquires its final globular conformation at the expense of the compaction ratio perpendicular to the axis, which decreases again significantly. If the thick filament is initially in contact with only one of the hemispherical caps or diffusion drives it close to one of these caps (snapshots at 35 and 40 ms in Figure 7e,f), then the cap behaves like an attractor and the DNA coil gets trapped therein. The points at $N = 2000$ in Figure 6 were obtained by averaging $\langle R_{\text{coil}} \rangle$, $\langle h_{\text{CM}} \rangle$, and $\langle \text{abs}(x_{\text{CM}}) \rangle$ for 75 ms (after equilibration) along these three trajectories. The abrupt displacement of the DNA coil towards a hemispherical cap upon increase of $N$ from 1750 to 2000 is very clear in Figure 6.

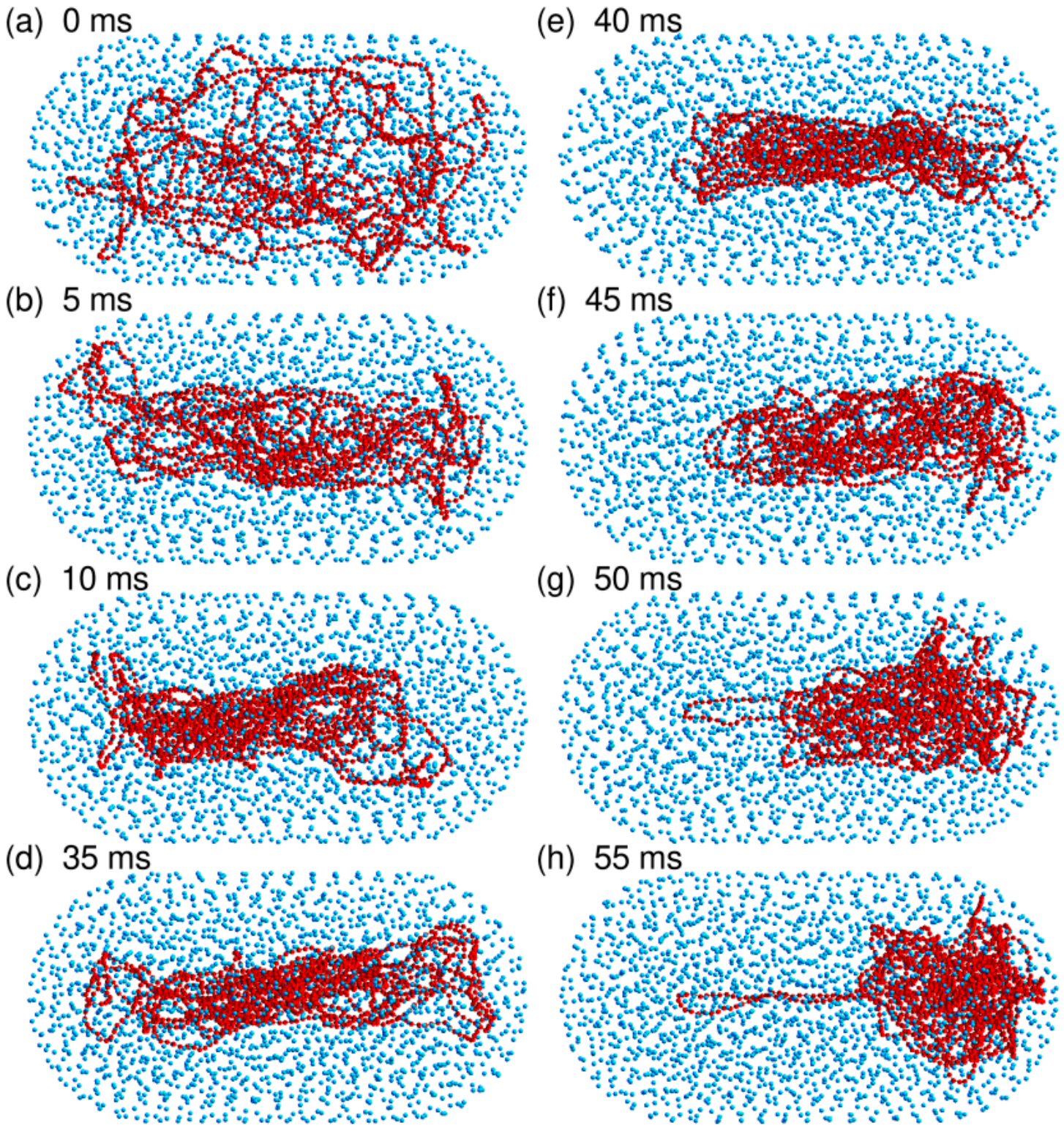

**Figure 7.** Sequence of snapshots at 0 ms (panel (**a**)), 5 ms (panel (**b**)), 10 ms (panel (**c**)), 35 ms (panel (**d**)), 40 ms (panel (**e**)), 45 ms (panel (**f**)), 50 ms (panel (**g**)), and 55 ms (panel (**h**)), extracted from a trajectory run with the capped confinement cylinder, $N = 2000$ crowders, and without repositioning the center of the confinement sphere on top of the center of mass of the DNA chain. This sequence illustrates the equilibration phase of the system, where the DNA chain compacts progressively and migrates towards one of the hemispherical caps. $N = 2000$ corresponds to an effective volume fraction of the crowders $\rho = 0.66$, close to the jamming threshold for solid spheres. DNA beads are colored in red and spherical crowders in cyan. Crowders are represented at $\frac{1}{4}$ of their actual radius, in order that the DNA chain may be seen through the layers of crowders. The capped confinement cylinder is not shown.

In contrast, the remaining five trajectories run with $N = 2000$ were transiently trapped in conformations similar to the one shown in Figure 2d. The DNA coil displays a dumbbell-like geometry composed of a thin filament stretching along the axis of the cylinder and connecting two globules anchored in each of the hemispherical caps. This conformation is only metastable and thermal fluctuations are sufficient to detach one of the globules from the corresponding hemispherical cap within a few hundreds of ms. The DNA chain subsequently undergoes the sequence of transformations shown in Figure 7.

Simulations performed with the rod-like confinement chamber therefore confirm the tentative conclusion drawn in the previous sub-section that the DNA coil abruptly acquires the localization properties of solid spheres close to the jamming threshold. This transition manifests itself through the preferential localization of the coil close to the regions of the wall with the largest curvature. In the simplest cases, this leads to a compact DNA globule filling one of the hemispherical caps of the confinement chamber. However, the attachment strength of a DNA coil to a spherical wall is large enough for the DNA chain to remain trapped for long times in a dumbbell conformation if thermal fluctuations do not perturb the symmetry of the coil during the initial steps of the compaction too much.

### *3.4. The Bacterial Nucleoid Does Not Behave Like A "Soft Sphere"*

In a recent work, Shew et al. investigated the localization of a large sphere in a cavity filled with smaller ones and suggested that the large sphere localizes preferentially at the periphery of the cavity when it is "rigid", but switches into the inner region when its "softness" increases [28]. Since the questions raised in their work and the present paper are clearly related, we performed several sets of simulations, in order to check whether their model is relevant to understanding the preferential localization of compact nucleoids inside bacterial cells.

The "soft sphere" model of Shew et al. [28] actually consists of a sphere, which has two different effective radii, namely an outer radius $b_S$ that governs the interactions with the bounding wall and an inner radius $b_S - \delta$ ($\delta \geq 0$) that governs the interactions with the small crowding spheres. Shew et al. showed that increasing $\delta$, that is allowing the small crowders to penetrate more and more deeply inside the outer radius $b_S$ of the large sphere, leads to the displacement of the average position of the large sphere towards the center of the spherical confinement chamber. The main issue with the simulations presented in Ref. [28] is the small size of the investigated system. Indeed, the outer radius of the large sphere, $b_S$, is approximately one half of the radius of the confinement sphere, $R_0$, which actually precludes a clear interpretation of the results. In order to clarify the results obtained with this model, we ran several sets of simulations along the same lines as in Ref. [28], but with a significantly larger $R_0/b_S$ ratio.

More precisely, the system investigated here consists of a large sphere of radius $b_S = 20$ nm enclosed in a confinement sphere of radius $R_0 = 125$ nm together with $N = 1499$ smaller crowding spheres of radius $b = 6.5$ nm. Crowder/crowder interactions obey the $V_{C/C}$ potential of Equation (4), while the interactions between the large sphere and the crowders are expressed in the form (Equation (10))

$$V_{S/C} = e_C^2 \sum_{K=1}^{N} H(\|\mathbf{R}_K - \mathbf{R}_S\| - b - b_S + \delta)\ , \tag{10}$$

where $\mathbf{R}_S$ denotes the position of the center of the large sphere, in order to account for the "softness" of the large sphere. Moreover, the repulsion exerted by the bounding wall was slightly modified according to (Equation (11))

$$V_{\text{wall}} = \zeta\left(\left(1 + \frac{D_S}{R_0 - b_S}\right)^6 - 1\right) + \zeta \sum_K \left(\left(1 + \frac{D_K}{R_0 - b}\right)^6 - 1\right), \tag{11}$$

where $D_S$ denotes the distance from the center of the large sphere to the bounding wall, so that the wall repels all the spheres as soon as their periphery (instead of their center) trespasses the wall. Effective volume fraction of the crowders is $\rho = 0.52$.

The first set of simulations was performed with $\delta = 0$, that is for a "rigid" large sphere. As illustrated in the inset of Figure 8, which shows the time evolution of the distance $r$ from the center of the large sphere to the center of the confinement sphere, the large sphere spends most of the time at rather well defined distances from the center of the confinement sphere and the transit time between two preferred shells is quite short. This is also clearly seen in the main plot of Figure 8 (dashed blue line), which shows the corresponding enrichment in the probability of finding the large sphere at distance $r$ from the center of the confinement sphere, $Q_S(r)$, obtained by averaging the computed density over 80 ms time windows from eight different trajectories. The most stable conformation, associated with the most intense peak at 105 nm, corresponds to the case where the periphery of the large sphere with radius $b_S = 20$ nm is in contact with the bounding wall (radius $R_0 = 125$ nm). The next peaks, of globally decreasing intensity, are separated by about 15.8 nm, close to $2(b + \Delta b) \approx 16.6$ nm. This indicates that the successive preferred shells correspond to conformations, where the periphery of the large sphere is separated from the bounding wall by an increasing number of layers of small crowding spheres. This result is in perfect agreement with previous ones dealing with similar systems [15–20].

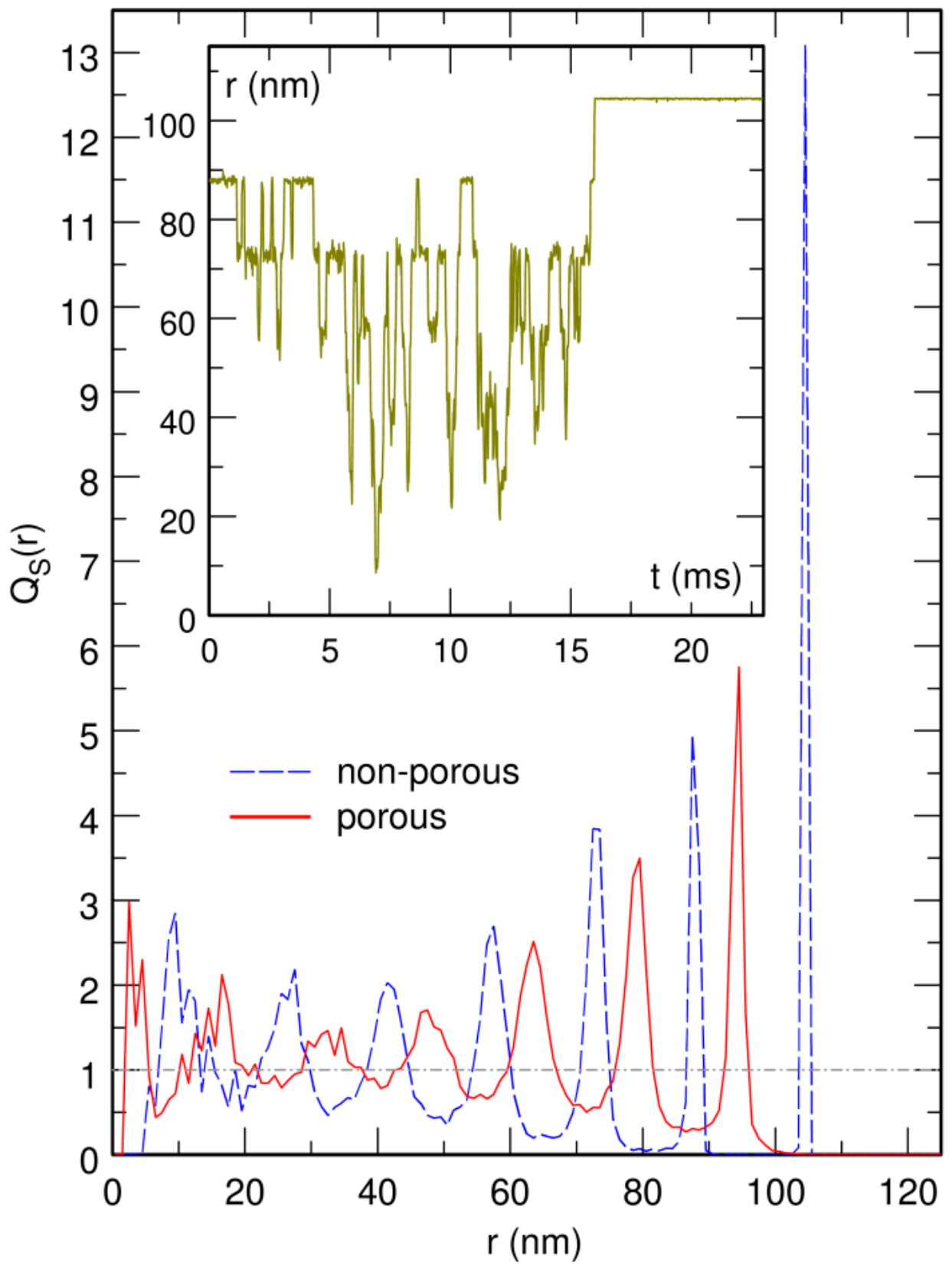

**Figure 8.** (**Inset**) Time evolution of $r$, the distance from the center of the large "rigid" sphere with radius $b_S = 20$ nm to the center of the confinement sphere. (**Main**) Plot of $Q_S(r)$, the enrichment in the probability of finding the large sphere at distance $r$ from the center of the confinement sphere, for the "rigid" case (dashed blue line) and the "soft" case (solid red line). $Q_S(r)$ is constant and equal to 1 (dot-dashed horizontal gray line) if the probability of finding the large sphere is uniform inside the confinement sphere. Time is expressed in units of ms and $r$ in units of nm. The effective volume fraction occupied by the $N = 1499$ crowders of radius $b = 6.5$ nm is $\rho = 0.52$.

A second set of simulations was then performed with $\delta = b$, that is, in the case where the small crowding spheres can penetrate into the large "soft" one over a distance $\delta$ equal to their radius $b$. The result is shown as a solid red line in the main plot of Figure 8. It is seen that the plot of $Q_S(r)$ consists, as for the "rigid" sphere, of a series of peaks separated by about 15.8 nm. The only difference is that the most stable conformation, associated with the most intense peak at 95 nm, corresponds to the case where the center of the large sphere is separated from the bounding wall by a distance of about $b_S + 2(b + \Delta b) - \delta \approx 30$ nm, which allows for a layer of small crowding spheres to intercalate between the inner periphery (radius $b_S - \delta$) of the large "soft" sphere and the bounding wall. This conformation is obviously favored by entropy with respect to the conformation where the outer periphery of the large sphere (radius $b_S$) is in contact with the bounding wall, the reason being that in this latter conformation, a crescent located between the inner periphery of the "soft" sphere (radius $b_S - \delta$) and the bounding wall remains empty, which ultimately reduces the total volume available for the small crowding spheres.

The results presented above therefore indicate that the "soft" sphere proposed by Shew et al. [28] is fundamentally a solid, in the sense that it localizes preferentially in a series of concentric shells, with the occupation probability being maximum in the shell closest to the bounding wall, just like a hard sphere does. Increasing the "softness" of the large sphere merely modifies the most stable conformation by selecting the one associated with maximum entropy. While this ultimately results in the "soft" sphere moving to some extent away from the bounding wall, it nevertheless does not lose its solid-state properties. In contrast, the circular chain investigated in the present work displays no particular affinity for the bounding wall up to crowder concentrations close to the jamming threshold, where it rather abruptly sticks to the wall. This phenomenon can consequently not be described as the transformation of the DNA coil from a "soft" sphere to a "rigid" one, according to the terminology of Ref. [28], although the DNA coil does acquire the localization properties of solid spheres close to the jamming threshold.

## 4. Discussion

The work reported in this paper elaborates on recent simulations, which add weight to the conjecture that the formation of the nucleoid may result from a segregative phase separation mechanism driven by the de-mixing of the DNA coil and non-binding macromolecules, presumably functional ribosomes [44,57,58]. In a previous work, the center of the confinement sphere was repositioned on top of the center of mass of the DNA chain after each integration step, in order that the results depend as little as possible on the interactions between the DNA chain and the bounding wall. The present work instead focuses on the preferential localization of the DNA coil inside the confinement chamber when the centering step is omitted. While it is a general belief that the nucleoid 'remains centrally located due to entropic repulsion from the cell wall' [87], the simulations reported here suggest that this statement is not valid at large crowder density, close to the jamming threshold, where the compaction ratio of the DNA coil increases abruptly [44,57,58]. Indeed, in this regime, the compact DNA coil localizes preferentially close to the bounding wall, and more precisely, close to the regions of the bounding wall with the largest curvature, as if it were a solid sphere [15–20]. Most importantly, the preferential localization of the DNA coil in the hemispherical caps of rod-like cells reported here is in striking contrast with experimental results, which show, unambiguously, that compact nucleoids are preferentially positioned at the center of mononucleoid cells and at one quarter and three quarters of the length of binucleoid cells [5,6,10,12–14,21–26]. The coarse-grained model therefore supports the hypothesis that the observed preferential localization of compact nucleoids at regular cell positions involves either some anchoring of the DNA molecule to the cell membrane or some dynamical localization mechanism.

At this point, it must be emphasized that, for the model discussed here, the effective volume fraction of crowders at the jamming threshold ($N = 2000$) is $\rho = 0.66$, but the volume fraction occupied by naked crowders, without taking electrostatic repulsion into account, is only $(4\pi N b^3)/(3\ V) \approx 0.32$. Consequently, the volume fraction, which is virtually accessible to water and smaller solutes, is of the order of 0.7, in perfect agreement with the 0.7 water volume fraction reported for *E. coli* [88]. Moreover, there is experimental evidence that the translational diffusion coefficient of macromolecules is about one order of magnitude smaller in bacterial cells than in water and about three times smaller than in eukaryotic cells [89], which suggests that the cytoplasm of bacteria is indeed close to jamming and that the largest crowder concentration discussed in the present paper is indeed relevant in vivo.

On the other hand, it must be admitted that it is far from being easy to figure out what the precise localization mechanism could be. Indeed, we are not aware of any system capable of inducing the regular localization of compact nucleoids through the periodic oscillation of the concentration of some chemical compounds, like the control exerted by the Min system on the position of the Z-ring [38]. Moreover, there are several well characterized examples of proteins that anchor the DNA to the poles of the cell [31–33], but we are aware of only one protein that anchors the DNA at mid-cell (and other regular positions), namely the Noc protein, which ensures that the septum does not form over the nucleoid of *B. subtilis* by occluding the division apparatus [37]. The point, however, is that it is not clear whether the compact nucleoid of *B. subtilis* localizes at mid-cell because the Noc protein is able to tether the DNA only at this very precise position, or the nucleoid would localize at mid-cell even in the absence of any Noc protein. Furthermore, most experimental results were obtained with *E. coli* [5,10,12–14,21,22,25,26], where the Noc protein is replaced by the SlmA protein, which provokes depletion of the cytokinetic ring protein FtsZ at mid-cell by mediating its binding to random positions along the DNA molecule, but does not anchor the DNA to the membrane [90,91].

Moreover, entropic demixing, which in the present model is responsible for the strong compaction of the DNA coil close to the jamming threshold, is rather sensitive to the parameters of the model, and in particular, to the detail of the various repulsion potentials [58]. In this regard, we note that no preferential localization of the DNA coil close to the hemispherical caps of cylindrical confinement chambers was observed in the simulations reported in Ref. [29]. While the reason therefore is probably that this work did not consider sufficiently large crowder concentrations, so that only mild compaction of the DNA chain was obtained, it cannot be fully excluded that the two models actually lead to different results. Generally speaking, further work is certainly needed to ascertain which kind of model is able to provide a reliable picture of the mechanisms that take place in real cells.

To conclude, it may be worth noting that the question investigated in the present work is rather reminiscent of the tight spatial regulation of the origin of replication (*oriC*) of slowly growing *E. coli*. Indeed, at the beginning of the cycle, *oriC* localizes at mid-cell, while the two daughter *oriC* move to quarter cell positions shortly after the beginning of replication and remain there till the two daughter cells separate [92,93]. The mechanism that allows for such precise positioning of *oriC* is also unknown, although early biochemical experiments suggest that the region of DNA around *oriC*, the terminus, and the site of ongoing replication, are able to associate with the membrane [94–96]. Owing to the obvious link between these two questions dealing with localization, it is not unreasonable to hope that progress along one of them will help to solve the other one.

**Funding:** This research received no external funding.

**Conflicts of Interest:** The author declares no conflict of interest.